\documentclass[pra,twocolumn,showpacs,preprintnumbers,amsmath,amssymb]{revtex4}
\usepackage{graphicx,subfigure}
\usepackage{dcolumn}
\usepackage{bm}
\usepackage{color}
\usepackage{soul} 

\begin{document}

\title{Fluorescence spectra of atoms in a phase-stabilized magneto-optical trap as an optical lattice}
\author{Seokchan Yoon}
\altaffiliation{Present Address: Department of Physics, Korea University, Seoul 136-701, Korea}
\author{Youngwoon Choi}
\altaffiliation{Present Address: School of Biomedical Engineering, Korea University, Seoul 136-701, Korea}
\author{Sungsam Kang}
\altaffiliation{Present Address: Department of Physics, Korea University, Seoul 136-701, Korea}
\author{Wook-Rae Kim}
\altaffiliation{Present Address: Samsung Electronics Co., Ltd., Suwon 443-803, Korea}
\author{Jung-Ryul Kim}
\author{Kyungwon An}
\email{kwan@phya.snu.ac.kr}
  \affiliation{Department of Physics and Astronomy, Seoul National University, Seoul 151-747, Korea}
 
\date{\today}

\begin{abstract}
We present a study on characteristics of a magneto-optical trap (MOT) as an optical lattice.
Fluorescence spectra of atoms trapped in a MOT with a passively phase-stabilized beam configuration have been measured by means of the photon-counting heterodyne spectroscopy.
We observe a narrow Rayleigh peak and well-resolved Raman sidebands in the fluorescence 
spectra which clearly show that the MOT itself behaves as a three-dimensional optical lattice. 
Optical-lattice-like properties of the phase-stabilized MOT such as vibrational frequencies and lineshapes of Rayleigh peak and Raman sidebands are investigated systematically for various trap conditions.
\end{abstract}

\maketitle

\section{Introduction}
A magneto-optical trap (MOT)~\cite{PhysRevLett.59.2631} has so far been used as a convenient tool to 
prepare samples of cold atoms for the first step for making ultra cold atomic ensembles.
Characteristics of a MOT itself has been extensively investigated, 
especially on the mechanisms of laser cooling and trapping.
However, studies with MOTs have been limited in some respects.
The atoms trapped in MOTs always interact with near-resonant laser fields 
and inhomogeneous magnetic fields, which result in relatively high temperature 
and undefined internal states of atomic ensembles.
Moreover, interference patterns and polarization states of six trapping laser beams 
are unstable and uncontrollable unless the phases of the laser beams are locked to each 
other.
Consequently, microscopic features of a MOT such as quantized atomic motion observed 
in optical molasses or lattices~\cite{PhysRevLett.65.33, PhysRevLett.69.49, PhysRevLett.70.410, Grynberg_PRL_1993, PhysRevA.55.R3987} 
have often been neglected.
The trapped atoms are then described as an ensemble of thermalized classical particles moving almost freely with a Gaussian velocity distribution in a large potential well.

Schadwinkel {\it et al.}~\cite{PhysRevA.61.013409} have experimentally shown that 
a MOT can create an optical-lattice structure with an intrinsically phase-stabilized 
beam configuration~\cite{OptComm.148.45} and the trapped atoms can be localized well in optical micropotentials.  
With sub-Doppler laser cooling~\cite{Dalibard:89}, the trapped atoms can be cooled 
down to micro-K temperature 
and localized well in light-induced micropotentials of the optical lattice formed by the MOT itself.
The behaviors of an atomic ensemble in a MOT which behaves as an optical lattice simultaneously are nontrivial.
It might pose some interesting questions: 
for example, one may ask if the localization of atoms in individual wells in an operating MOT would suppress 
light-induced intra-trap cold collisions~\cite{cold_collision1, cold_collision2, cold_collision3} 
like as in optical lattices~\cite{PhysRevLett.80.480}.

There have been several nonlinear spectroscopic researches showing narrow Raman 
resonances having sub-natural linewidth in operating 
MOTs~\cite{Europhyslett.15.149, PhysRevLett.66.3245, PhysRevA.71.013401, LaserPhysLett.7.321}.
These experiments were carried out by means of pump-probe Raman transmission spectroscopy. 
There also have been several researches that measure the fluorescence spectrum of the atomic ensemble 
confined in optical molasses by means of optical heterodyne technique~\cite{PhysRevLett.65.33, PhysRevLett.69.49} 
and the second-order correlation function of the fluorescence light emitted from the atoms 
in optical molasses~\cite{PhysRevA.53.3469, PhysRevA.68.013411, Nakayama:10}.
Recently, near-resonant fluorescence spectrum of a single atom~\cite{NanoLett.11.729} 
confined in a three-dimensional (3D) optical lattice formed by a phase-stabilized MOT 
have been measured by means of the heterodyne photon-counting-based 
second-order correlation spectroscopy (PCSOCS)~\cite{Hong:06}.
Here the observed single-atom spectrum showing a strong Dicke narrowing~\cite{Dicke}
and motional Raman sidebands reveals the atom in a MOT can be bounded tightly in a light-shift micropotential.
The Mollow triplet spectrum of a few atoms near the optical nanofiber~\cite{Das:10} 
has been measured by the heterodyne PCSOCS.

A spectrum of scattered light from cold atoms gives much information on 
internal/external states of individual atoms and also on properties of an atomic ensemble.
The spectrum of individual atoms trapped in a phase-stabilized MOT is characterized by the intrinsic properties 
of the optical lattice rather than by the properties of the MOT such as vibrational frequencies and linewidths of Rayleigh peak and Raman sidebands.
On the other hand, the spectrum also contains information on characteristics of an atomic ensemble  
in the MOT such as temperature or dynamics of cold atomic collisions. 
In a dense atomic gas, for example, the atoms experience radiation trapping~\cite{radiation_trapping1}, 
reabsorption of emitted photons from other atoms, which would increase the temperature and
results in an additional homogeneous broadening in the spectrum. 
Cold collisions between trapped atoms would also introduce an additional spectral broadening.
In addition, fluorescence spectroscopy allows ``nondestructive'' measurements on cold atomic ensembles
in contrast to the pump-probe spectroscopy.
However, a {\em systematic} study on fluorescence spectra of an atomic ensemble in an operating MOT as an optical lattice
has not yet been reported.
In this paper, motivated by these considerations, we present a comprehensive fluorescence spectroscopic 
study of a MOT as an optical lattice by means of the heterodyne PCSOCS.

This paper is organized as follows. In Sec.~\ref{sec:latticeMOT}, 
a beam configuration of our passively phase-stabilized MOT and the structure of 
a 3D optical-lattice in a MOT are discussed.
In Sec.~\ref{sec:setup}, a detailed description of experimental scheme for photon-counting 
heterodyne spectroscopy is presented.
We emphasize that our photon-counting heterodyne technique is a non-destructive, almost real-time measurement of
the spectrum (thus the temperature information) even for a small number of cold atoms, $N\sim$100. 
In Sec.~\ref{sec:results}, 
the measured atomic fluorescence spectra from the MOT in a phase-stabilized configuration
and also in a standard retroreflected six-beam configuration are shown and discussed.
The optical lattice properties such as Raman vibration frequencies have been 
investigated systematically for various trap parameters.
The results are compared with the theoretical calculations of the energy-band of the optical lattices.
We also discuss spectral broadening in the fluorescence spectra due to many-atom effects
for dense atomic ensembles.
Finally, we summarize our work in Sec.~\ref{sec:conclusion}.

\section{MOT as an Optical Lattice}
\label{sec:latticeMOT}

In a stable optical lattice, phase fluctuations in the laser beams forming the lattice should 
just induce an overall translation of the lattice structure, not affecting 
the topography of micropotential significantly.
It is known that one has to use only four beams in order to construct such a stable lattice 
in 3D~\cite{Grynberg_PRL_1993}. 
For a standard MOT in $\sigma^+ - \sigma^-$ beam configuration, however, six trapping laser beams are
usually used.
The relative time phases of individual trapping laser beams change independently and randomly in time 
due to mechanical vibrations or thermal drifts of optics.
It is thus difficult to establish a stable optical lattice with its own trapping laser fields in a typical MOT configuration.
A stable optical lattice, indeed, can be established in a MOT by using a phase-stabilized-beam configuration, 
{\it i.e.}, by introducing the two-beam configuration which has been demonstrated 
by Rauscheneutel {\it et al.}~in Ref.~\cite{OptComm.148.45}.
In this configuration, effectively one standing wave is formed by two phase-related beams and it is 
folded and intersecting with itself twice.
The time phases of the trapping beams are related to each other and thus the interference pattern 
is intrinsically stable.
The beam configuration for the phase-stabilized MOT used in our experiments is described 
in Ref.~\cite{arXiv} in more detail. 

The intensity of total electric field of the MOT with proper axes 
and origin is then given by 
$I({\bf r}) = 12I_0 f({\bf r})$, 
where the $I_0$ 
is the intensity of a single beam and $f({\bf r})$ 
having a value between 0 to 1
is a geometrical structure function of the interference pattern given by
$f({\bf r}) = 1/2 + (\cos{kx}\sin{ky} + \sin{kx}\sin{kz} - \cos{ky}\cos{kz})/3$.
The local polarization of the total electric field is linear everywhere but its orientation varies in space.
The optical dipole potential for a multi-level atom in a hyperfine 
Zeeman ground sublevel $|F_g, m\rangle$ interacting with a linearly polarized electric 
field of the MOT is given by
\begin{equation}
  U_{m}({\bf r}) = \frac{\hbar\varDelta}{2} \log \left[1+C^2_m Sf({\bf r})\right], 
\label{eq:dipole_potential}
\end{equation}
where $\varDelta$ is the laser detuning, $C_m$ is the normalized Clebsch-Gordan coefficient for $\pi$ transition 
($|F, m\rangle \leftrightarrow |F', m\rangle$),
and $S$ is the detuning-dependent saturation parameter givin by 
\begin{equation}
S =\frac{12I_0/I_s}{1+4\varDelta^2/\Gamma^2} \nonumber
\end{equation}
with the maximum intensity of $12I_0/I_s$.
The saturation intensity $I_s$ is 1.67 mW/cm$^2$ for $F=3 \rightarrow F'=4$ 
cycling transition of $^{85}$Rb atom with circularly ($\sigma^{\pm}$) polarized light
and $\Gamma/(2\pi)=6$ MHz is the full-width-at-half-maximum (FWHM) natural linewidth of the transition.

\begin{figure}
\includegraphics[width=3.4in]{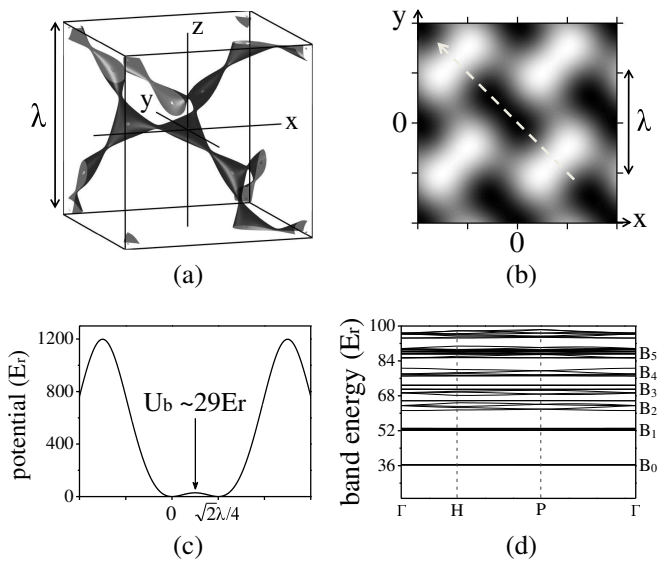}
\caption[The optical lattice in a MOT]
{
(a) 3D equipotential surface for $U({\bf r})=30 E_r$ in a unit cell of the optical lattice potential in a MOT
with the origin at the position of minimum potential.
(b) 2D contour map on $x-y$ plane.
(c) potential along the white dashed line shown in (b) connecting two adjacent potential minima.
The potential minima are connected through a shallow double-well barrier $U_b$. 
(d) Energy band structures of 3D optical lattice in a MOT. Only the ground and first excited 
bands (B0 and B1) are tightly bound in double-well micropotentials under the the above trap condition. 
Here $E_r$ is the photon recoil energy defined in the text. 
The trap parameters used in the calculation are $\varDelta=-4\Gamma$ and $S=1.0$.
}
\label{fig:lattice_structure}
\end{figure}

The 3D structure of the lattice and the contour map for a specific two-dimensional (2D) plane 
for the optical potential are depicted in Fig.~\ref{fig:lattice_structure}.
The lattice forms a body-centered-cubic structure containing eight local minima in a unit cell.
Each minimum has three nearest neighboring minima and these minima are connected 
to each other through a relatively shallower potential barrier with a height of $U_b$.
The potential height in the direction perpendicular to the plane defined by three nearest-neighbor 
minima is much more higher than the barrier height $U_b$.
Hence, the structure of the optical lattice looks like an array of chained 
anisotropic double-well potentials in 3D space.
For more details see Ref.~\cite{NanoLett.11.729}.

The intrinsic characteristic of the optical lattice in a MOT with the double-well 
barrier $U_b$ is determined only by $f({\bf r})$, the geometrical structure of the interference pattern.
The height $U_b$ of the double-well barrier between two adjacent potential minima is given by
\begin{equation}
U_{b} =
\frac{\hbar\varDelta}{2} \log \left[1+(-1/2+\sqrt{2}/3) \frac{C^2_mS}{1+C^2_mS}\right], \\
\label{eq:Ub}
\end{equation}  
where $C^2_{m=0}=4/7$ and $C^2_{m=\pm1}=15/28$ for $|F=3, m\rangle \rightarrow |F'=4, m\rangle$ 
transition of a rubidium atom.
The barrier $U_b$ does not increase linearly as the intensity increases 
at a fixed detuning and is saturated to a finite level when intensity increases to infinity.
The maximum height of the barrier is then $U_{b\_max} \approx 23 (|\varDelta|/\Gamma) E_r$, 
where $E_r$ is a recoil kinetic energy of an atom. 
For example, when the detuning of the trap laser is $-4\Gamma$, 
the barrier height $U_b$ is limited to $92E_r$ corresponds to $34 ~\mu$K 
in one-dimensional temperature for a rubidium atom with a recoil energy $E_r/h=3.86$ kHz.
Since the $U_b$ is too shallow for a usual trapping condition with a detuning of 
$\Delta=-\Gamma \sim -6\Gamma$ and a intensity $I_0= I_s \sim 3 I_s$, 
only the motional ground and first excited bands (B0 and B1) can be tightly bound three dimensionally 
as shown in a diagram of energy-band structure depicted in Fig.~\ref{fig:lattice_structure}(d).
The atoms populated in higher excited bands can move relatively freely through the shallow channels
but are still confined in other directions.

\section{Experimental Setup}
\label{sec:setup}

\subsection{Setup for MOT}
In this experiment $^{85}$Rb atoms are trapped in a vapor-loaded MOT.
Only two phase-related beams are used in order to allow a stable 3D optical lattice in the MOT.  
The frequency of the trap laser is locked to D2 transition line (5S$_{1/2}$, $F=3$ 
$\rightarrow$ 5P$_{3/2}$, $F'=4$) of $^{85}$Rb atom by means of frequency 
modulation technique and is typically detuned to $-1\Gamma \sim -6\Gamma$ from the atomic resonance. 
A Rubidium dispenser is mounted about 3 cm below the MOT center.
Only the number of trapped atoms in a MOT can be varied by changing the driving current on the dispenser 
while the other trap parameters are all fixed. 
When a magnetic field gradient is set to 60 G/cm, typically about $10^2 \sim 10^5$ 
atoms are loaded to the MOT and the size of the atom cloud is about 50 $\mu$m in the 1/e radius.

\subsection{Setup for heterodyne measurement}
\begin{figure}
\includegraphics[width=3.4in]{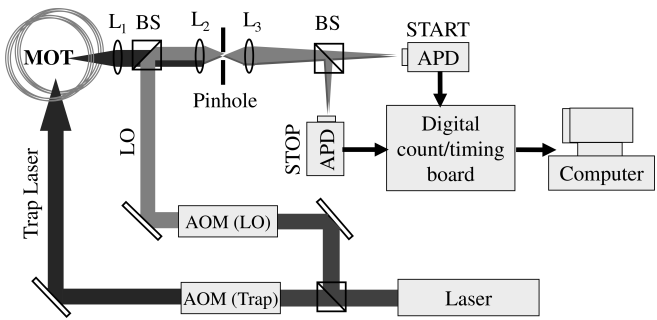}
\caption[Experimental setup]{Schematic diagram of the experimental setup 
for the heterodyne PCSOCS. 
AOM Acousto-optic modulator, LO local oscillator, BS beam splitter, L lens, APD avalanche photo diode.
}
\label{fig:heterodyne_setup}
\end{figure}

The setup for the spectrum measurement based on the heterodyne PCSOCS 
is shown in Fig.~\ref{fig:heterodyne_setup}.
A local oscillator (LO) is derived from the same laser as the trapping laser
and the frequency is shifted by an acousto-optic modulator (AOM) and actively locked to 10 MHz apart from the trapping laser frequency.
The fluorescence emitted from the atoms trapped in a MOT 
is collected by an objective lens $L_1$ ($f_1$=25 mm, numerical aperture of 0.26) and combined with LO by a beam splitter (BS). The combined lights are then spatially filtered in order to match the spatial mode.
Finally it is split into two beams and detected by two avalanche photo diodes (APDs) operating in photon counting mode.
A detection volume is about 15 $\mu$m in diameter and the mean photon count rate of the
fluorescence light was typically about $10^5 \sim 10^6$ cps for each APD.
A background count rate is lower than 100 cps, which is negligibly small compare to the fluorescence signal.
The photon pulses detected by two APDs are fed into two counter/timing boards 
(NI-6602, National Instruments) as  START and  STOP pulses, respectively.
By accumulating a histogram of the number of stop photons arriving at 
time separation of $\tau=t_{\rm stop}-t_{\rm start}$, the intensity correlation function 
$g^{(2)}(\tau)$ is obtained.
Since this $g^{(2)}(\tau)$ contains the first-order correlation function of the fluorescence \cite{Hong:06},
the fluorescence spectrum can be obtained by performing a digital fast Fourier transform 
of $g^{(2)}(\tau)$ and displayed on a computer monitor almost in real time.
In this setup, we typically use a bin time of 12.5 ns and a span time of 10 ms, which correspond to 
0.1 kHz frequency resolution and 40 MHz maximum span frequency, respectively.
The spectral resolution of our detection system is measured to be 1.4 kHz which is mainly determined by the phase fluctuations of the rf signal driving the AOM used for frequency shift of LO laser.

\section{Results} \label{sec:results}

\subsection{Phase-stabilized MOT}

\subsubsection{Characteristic fluorescence spectrum}
\begin{figure}
\includegraphics[width=3.4in]{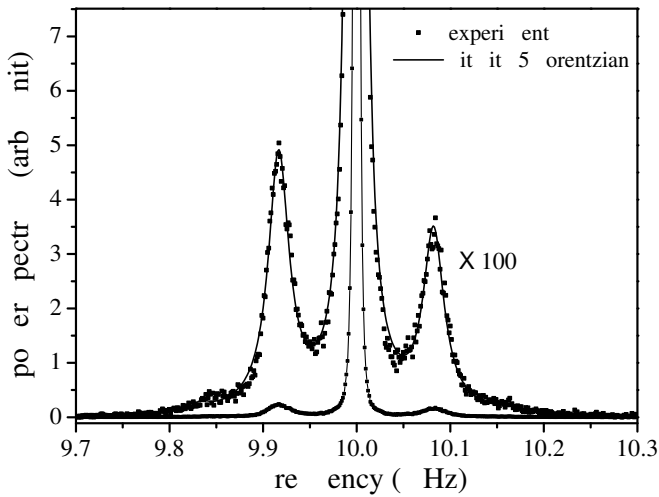}
\caption[A characteristic fluorescence spectrum of the atoms trapped in a MOT in 
the phase-stabilized beam configuration]
{A chararcteristic fluorescence spectrum of the atoms trapped in a MOT in the phase-stabilized beam configuration.
The detuning is $-3\Gamma$ and saturation parameter $S$ is about 1.
An elastic Rayleigh peak at the center and four Raman sidebands are clearly shown in the spectrum. 
The solid line is a fit with five Lorentzians. 
}
\label{fig:two_beam_spectrum}
\end{figure}
A characteristic fluorescence spectrum with the phase-stabilized MOT is shown in Fig.~\ref{fig:two_beam_spectrum}.
Typically a small number of atoms ($N\sim 100-1000$) are loaded in the MOT 
with an atomic number density $n\sim10^9$/cm$^3$.
In contrast to a usual MOT where one might expect a Doppler broadened spectrum, 
the spectrum in Fig.~\ref{fig:two_beam_spectrum} does not show Doppler broadening. 
Instead, it shows a sharp central peak and small sidebands.
We interpret that the central peak is due to elastic Rayleigh scattering or photon emission without 
changing both the vibrational band $B_n$ and Zeeman $m$-sublevel of ground state.
The first two sidebands at positions $\pm\nu_{\rm osc}$ with respect to the central peak
correspond to Raman transitions between the two lowest vibrational bands, $B_0$ and $B_1$.
On the outside of them, two more broader and much smaller sidebands are barely seen. 
We attribute them to the transition between the first and third vibrational bands, $B_0$ and $B_2$.
The spectrum is well fit by five Lorentzians for the Rayleigh peak and the four Raman sidebands without considering Doppler broadening.
The disappearance of Doppler broadening in the spectrum is associated with Dicke narrowing 
of strongly confined atoms in space, {\it i.e.}, 
a strong 3D confinement of atoms inside the shallow lattice potentials of the MOT.

We consider here that the local polarization of the total electric filed of the MOT's trapping lasers is linear everywhere.
Due to the optical pumping, the atoms are mostly populated in $m$=0 and $m$=$\pm1$ Zeeman sublevels 
of the ground state with relative populations, $w_0$=0.38 and $w_{\pm1}$=0.24~\cite{PhysRevA.61.013409, PhysRevA.71.013401}.
The different Clebsch-Gordan coefficients $C^2_m$ will induce different light-shift 
potential depths for different $|m|$ sublevels~\cite{LaserPhysLett.7.321}:
the difference of potential well depths between $m=0$ and $|m|=1$ ground-state Zeeman sublevels 
is expected to be in the range of 200$\sim$400 kHz under the typical MOT conditions. 
Thus, it is expected to observe additional peaks due to the electronic 
spontaneous Raman transitions ($|m,B_n\rangle\rightarrow|m\pm1,B_n\rangle$) in the vicinity of several hundred kHz
as shown in probe transmission spectra~\cite{PhysRevA.61.013409, PhysRevA.71.013401, LaserPhysLett.7.321}.
In the fluorescence spectra, however, we could not find any observable peaks 
to be considered as such electronic Raman transitions between different Zeeman $m$ sublevels.
A possible reason would be an inhomogeneous magnetic field near the center of the magnetic quadrupole of MOT.
Due to the finite size of atomic cloud, the atoms experience different Zeeman shifts for $m\neq0$ sublevels at different locations.
Moreover, the atomic transition by the locally linearly polarized trapping laser cannot be regarded as a pure $\pi$ transition in the presence of arbitrary oriented magnetic fields. 
All of these would induce position-dependent light shits and inhomogeneous broadening in $|\Delta m|=1$ Raman transition spectra.

\subsubsection{Vibrational frequency}

\begin{figure}
\includegraphics[width=3.4in]{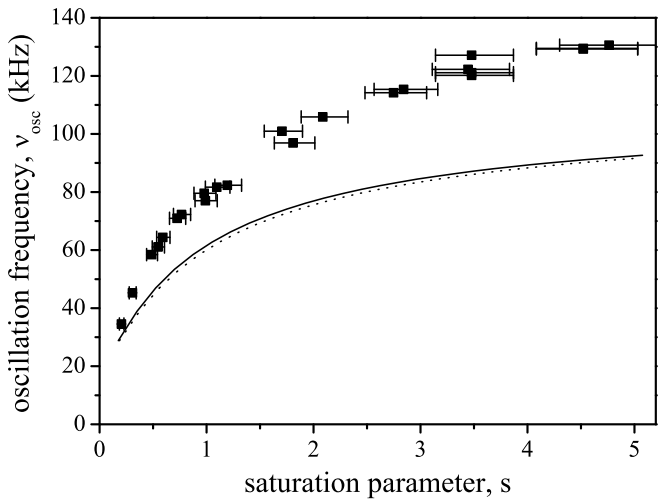}
\caption[Oscillation frequency as a function of saturation parameter]
{
Oscillation frequency $\nu_{\rm osc}$ as a function of saturation parameter, 
$S=12(I_0/I_s)/\left[1+4\varDelta^2/\Gamma^2\right]$. 
The trap-laser intensity $I_0$ is increased at fixed detuning of $\varDelta=-4\Gamma$.
Squares are experimental measurements and error bars indicate about 10\% uncertainty 
in intensity measurements.
Solid and dotted lines are calculated energy gaps between ground (B0) and 
first-excited (B1) states for $m=0$ and $|m|=1$ Zeeman sublevels, respectively.
}
\label{fig:osc_vs_parameter}
\end{figure}

The vibrational frequency $\nu_{\rm osc}$ corresponding to the peak position of 
the first sidebands is measured as a function of the saturation parameter $S$ by varying the intensity of the trapping laser at a fixed detuning of $\varDelta=-4\Gamma$. 
The measurements are compared with the band structure calculations in Fig.~\ref{fig:osc_vs_parameter}.
From the band structures calculated for the mostly populated atomic states $m=0$ and $\pm1$, 
the first energy gap between the ground ($B_0$) and first-excited ($B_1$) 
bands are obtained and plotted in solid and dotted lines, respectively. 
The shapes of the potentials for $m=0$ and $|m|=1$ are not significantly different 
and thus the vibrational frequencies corresponding to $|m=0,B_0\rangle \leftrightarrow |m=0,B_1 \rangle$ and 
$|m=\pm1,B_0\rangle \leftrightarrow |m=\pm1, B_1 \rangle$ transitions can not be distinguished
clearly from the observed spectra.

Both the theoretical and measured oscillation frequencies $\nu_{\rm osc}$ clearly show saturation 
trends with the saturation parameter $S$.
We confirmed that the square of the vibrational frequency $\nu^2_{\rm osc}$ is 
linearly proportional to the shallow barrier $U_b\propto C^2_mS/(1+C^2_mS)$ rather than the maximum potential depth, 
$U_{\rm max}\propto\log{(1 + C_m^2 S)}$ over a large range of $S$.
As we discussed in Sec.~\ref{sec:latticeMOT}, this saturation trend in the intensity dependence of $\nu_{\rm osc}$ can be interpreted as the saturation effect of $U_{b}$.  
It is interesting to note that the optical lattice formed by the MOT's field is characterized 
by the height of the shallow barrier $U_b$ rather than $U_{\rm max}$.

It is noted that the measured oscillation frequencies are about 1.4 times higher than those predicted by calculations, 
which means that the actual height of $U_b$ is roughly two times higher than the predicted $U_b$ in Eq.~(\ref{eq:Ub}).
We are not able to clearly explain the discrepancy between the measurements and the calculations.
It might be attributed to various experimental imperfections such as intensity imbalance, imperfect polarization or misalignment of the trap lasers, 
any of which can induce an unexpected distortion in the special interference pattern $f(\bf r)$, especially 
in the {\em fine microstructure} of the double-well potentials.
However, such imperfection is unlikely since in the experiments the trapping beams and position of the atom cloud were adjusted 
for the best lattice formation by monitoring changes in the spectral line shapes in real time.
The peak intensity of the Gaussian-shaped trapping beam is used to calibrate saturation parameter $S$.
It has been checked that the calibrated saturation parameter used in Fig.~\ref{fig:osc_vs_parameter} is 
in good agreement with that obtained from the saturation curve of single-atom fluorescence~\cite{yoon}.

\subsubsection{Spectral linewidth}

\begin{figure}
\includegraphics[width=3.4in]{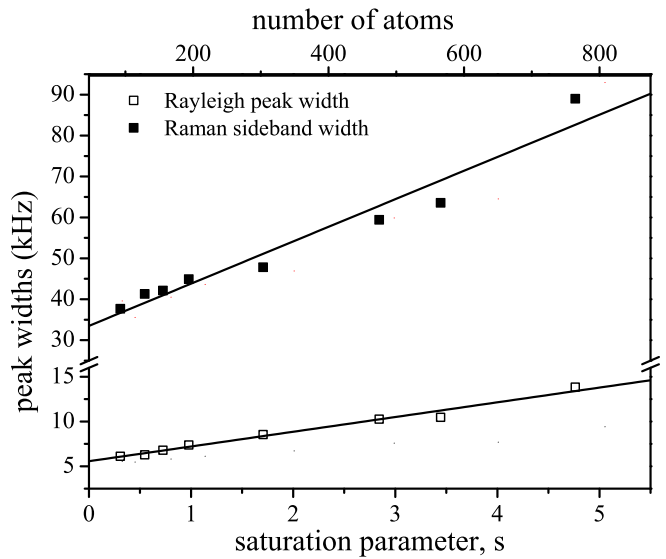}
\caption[Linewidths of Rayleigh peaks and first-order Raman sidebands as 
a function of saturation parameter.]
{Linewidths of Rayleigh peaks and first-order Raman sidebands as 
a function of saturation parameter. 
The open and solid squares represent the measured linewidth of the Rayleigh and Raman sidebands, 
respectively.
The solid lines are linear fits.
The trap-laser intensity $I_0$ is changed at fixed detuning of $\varDelta=-4\Gamma$.
The number of trapped atoms increases linearly with the intensity 
because of increasing of capture ability of the MOT.
}
\label{fig:two_beam_width}
\end{figure}

The linewidths of the Rayleigh peak and the Raman sidebands are shown 
as a function of saturation parameter $S$ in Fig.~\ref{fig:two_beam_width}.
Both the Rayleigh and Raman linewidths increase as $S$ increases.
There are several possible effects affecting the lineshape of the fluorescence spectrum.

The Rayleigh linewidth is related to the finite lifetimes of the vibrational states~\cite{PhysRevA.46.7060}.
In the fluorescence spectrum of a single atom in the MOT of Ref.~\cite{NanoLett.11.729}, 
it was observed that the optical pumping into $F=2$ hyperfine ground state, 
which is proportional to the atomic excited-state population or equivalently the trap-laser intensity, 
gives the main spectral broadening of the Rayleigh peak.
The tunneling rate of the bound atom from one lattice site to nearest neighbors gives 
additional broadening in the spectrum, but it is relatively small  
compare to the optical pumping at a large $S$, {\it i.e.}, a deep potential well. 
In a deep harmonic potential limit, the tunneling rates of the atoms in the deepest bound states would decrease 
as the potential depth increases and thus the Rayleigh linewidth is inversely proportional to $S$.
The measured linewidth of the Rayleigh peak in a single-atom spectrum in Ref.~\cite{NanoLett.11.729} was 
varied only in a range of $3\sim5$ kHz under various trap conditions due to the above broadening effects.

In Fig.~\ref{fig:two_beam_width}, however, the linewidth of Rayleigh peak is larger, 
varying linearly from 6 to 14 kHz as $S$ is increased and consequently as the number (density) of trapped atoms 
is varied from $\sim$100 ($3\times10^9$/cm$^3$) to $\sim$800 ($6\times10^9$/cm$^3$).
Nonetheless, the linear fit of the observed Rayleigh linewidth has a finite intercept of 5.6 kHz, 
which is quite comparable to the linewidth of a single-atom spectrum~\cite{NanoLett.11.729}.
The increased widths of the Rayleigh peak and the Raman sidebands observed in this measurement must be interpreted 
as extra homogeneous line broadenings due to many-atom effects such as atomic density-induced radiation trapping
and intra-trap cold collisions.
Due to the radiation trapping, a large number of atoms in a MOT usually give rise to a high atomic temperature. 
The hotter atoms become more populated in high-lying vibrational levels, which lead to higher tunneling rate 
of the atoms between adjacent potential minima through the shallow potential well $U_b$ and consequently 
give broader linewidth of the Rayleigh line.

The main cause of the broadening in the Raman sidebands is evidently the high temperature of the atoms 
in the fully anharmonic potential well of the MOT as in Ref.~\cite{PhysRevA.61.023410}.
Consequently, the larger atomic populations in high-lying vibrational levels at a large saturation parameter $S$ 
result in asymmetric Lorentzian lineshapes of the Raman sidebands and broadening of the whole spectrum.
We attribute such frequency shifts, as well as broadenings and asymmetric lineshapes of the Raman 
sidebands to the increase of atomic temperature in a strongly anharmonic 
band structure [as shown in Fig.~\ref{fig:lattice_structure}(d)].

\subsubsection{Spectrum vs.~density}

It is known that the effects of the radiation trapping become observable only in 
dense atomic gases with densities of $10^{10} - 10^{11}$/cm$^3$ in general.
It has been shown that the coherence properties of the scattered light measured 
by intensity correlation function can be extremely sensitive to the 
presence of the radiation trapping even for a dilute atomic gas with densities of $10^{8} - 10^{9}$/cm$^3$~\cite{PhysRevA.68.013411}. 

\begin{figure}
\includegraphics[width=3.4in]{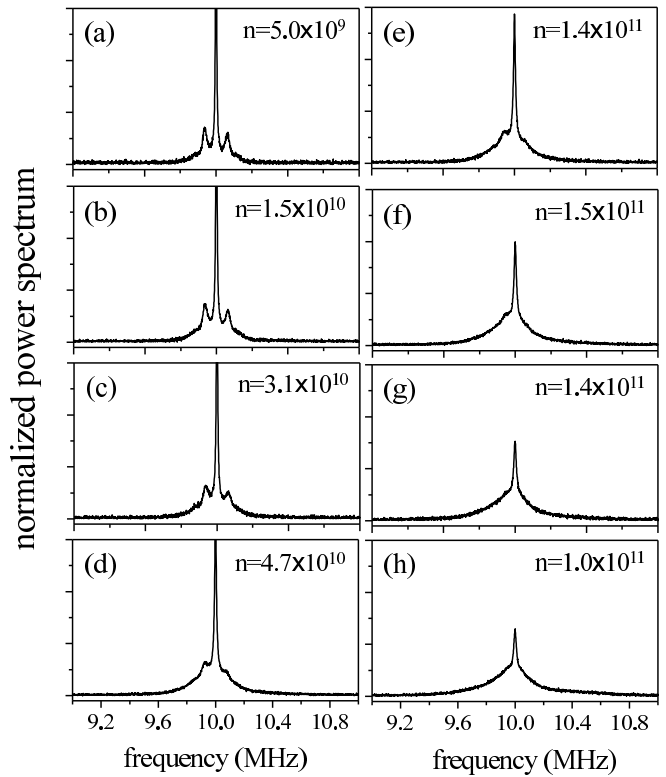}
\caption[Fluorescence spectrum vs. atomic density]
{Fluorescence spectrum vs.~number density $n$ of atom cloud. 
The current on the Rubidium dispenser is varied from 2.6 A for (a) to 4.5 A for (h). 
The atomic density is obtained from the size and the number of trapped atoms which 
are measured from the fluorescence image of atom cloud taken by a CCD camera. 
The numbers of trapped atoms are calibrated with respect to the independent-obtained 
fluorescence counts of single trapped atom.
}
\label{fig:spectrum_vs_density}
\end{figure}

In order to demonstrate how the spectral lineshape can be affected by 
the number density of atomic cloud, 
the fluorescence spectra were measured for various numbers of trapped atoms.
Only the number of trapped atoms is varied by changing the current applied on 
the Rubidium getter dispenser, while other trapping parameters are all fixed
($\varDelta=-4\Gamma$, $S=0.5$).
The density of atomic cloud is varied from $n\sim5.0\times10^{9}$/cm$^3$ to $1.5\times10^{11}$/cm$^3$,
as the number of atoms is being increased from about $100$ to $3.0\times10^{5}$.
The resulting spectra are shown in Fig.~\ref{fig:spectrum_vs_density}.
The widths of Rayleigh and Raman peaks become broader as the atomic density increases.

It is interesting that the spectral shape starts to change substantially when the density is increased
beyond $5\times10^{10}$/cm$^{3}$: 
The Rayleigh peak and Raman sidebands gradually disappear and a much broader Gaussian-shaped 
Doppler-broadened spectrum emerges.
We interpret this as a result of increase in atomic temperature due to the radiation trapping.
The increase in atomic kinetic energy leads to changing the atomic motional state 
from bound states in double-wells to upper continuum states. 
In fact, we have confirmed that the spectrum is changed in the same way  
when the atoms are heated up by shining a resonant laser beam directly to the atom cloud
even at a low atomic density. 

As the getter current increases linearly, the density and number of trapped atoms increase 
to their maxima $n=1.5\times10^{11}$/cm$^{3}$ and $N=2.4\times10^5$, 
respectively, and then start to decrease, but the spectrum still continues to become broader as shown 
in Fig.~\ref{fig:spectrum_vs_density} (g) and (h). 
These behaviors are due to the background gas collisions: 
the density of background Rubidium atoms increases with the getter current 
and thus induces collisions with the trapped atoms, 
thereby decreasing the number of trapped atoms and increasing the collisional broadening.
In this regime, the size of atom cloud increases without increasing in the atomic density. 

For a dilute atom cloud with a density of $10^8 - 10^9$/cm$^{3}$ in our phase-stabilized MOT,
only a small fraction of the total number of the lattice sites 
($8/\lambda^3\sim 10^{13}$ sites/cm$^3$) is occupied by the atoms. 
One of the interesting feature in this situation with a small filling factor is that the tightly 
bound atoms in the lattice sites would collide rarely with each others.
It is expected that light-induced two-atom collisions~\cite{cold_collision2,cold_collision3} 
are greatly suppressed.

\begin{figure}
\includegraphics[width=3.4in]{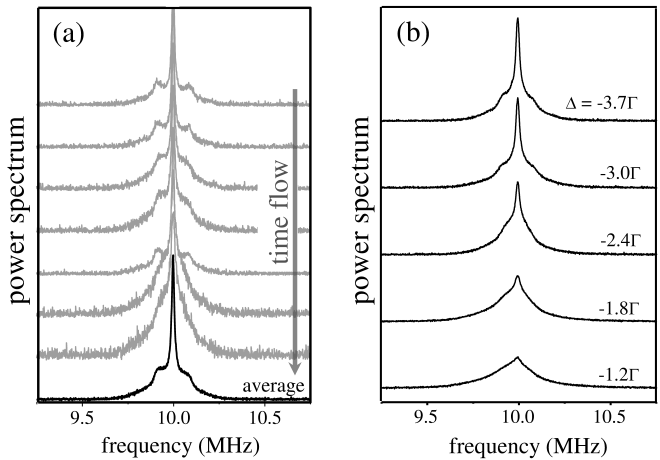}
\caption{Fluorescence spectra for a MOT in a typical retroreflected six-beam configuration. 
(a) The spectra in gray color are recorded for every 30 seconds and showed sequentially from top to bottom.
The bottom one is a total accumulated spectrum of all spectra.
(b) The spectra are measured under various detuning conditions.
Each spectrum is accumulated for about 10 to 20 minutes.
}
\label{fig:spectrum_vs_detuning}
\end{figure}

On the contrary, as the number (density) of trapped atoms are increased ($>10^{10}$/cm$^3$), 
the mean kinetic energy of the atoms becomes higher than the double-well barrier $U_b$ due to the 
radiation trapping and they start to move freely over the barrier.
However, the atoms are still well confined in effective 2D space because of the higher potential 
depth along other directions.
In this case the effective trap volume should be introduced as the volume 
in which the atoms can move freely. 
This volume is typically about 50 times 
smaller than the volume defined by the size of atomic cloud in an ordinary MOT 
without the lattice structure.
The density of atoms in terms of the effective volume has already reached the maximum ($\sim10^{13}$/cm$^3$) 
in Fig.~\ref{fig:spectrum_vs_density}(f) and cannot be increased further due to 
the light-induced intra-trap cold collisional loss. A substantial increase in two-atom collisions is expected there. 

\subsection{Fluorescence spectra in a standard  MOT}

We have also measured fluorescence spectra of trapped atoms in a MOT in a standard 
retroreflected six-beam configuration, 
where three trapping laser beams propagating along orthogonal axes are retro-reflected. 
The effects of phase fluctuations between the trapping beams on the fluorescence spectrum are shown in Fig.~\ref{fig:spectrum_vs_detuning}(a).
The each spectrum is recorded sequentially in every 30 seconds and shown in chronological order from the top.
The spectrum at the bottom is a total accumulated spectrum of all recorded spectra.
Strong Dicke narrowing and even Raman sidebands are clearly shown in some spectra, 
but Doppler-broadened distribution is dominant in most cases.
In this case, small changes in phase relation between the trap beams would generate significant 
changes in topography of light-induced micropotentials. 
Especially, the structures of shallow micropotentials are easly destroyed 
by small mechanical vibrations in optics.
The mechanical vibration is usually slow compare to the timescale of the atomic vibrational 
motion in micropotentials ($\nu_{\rm osc}\sim100$ kHz).

\begin{figure}
\includegraphics[width=3.4in]{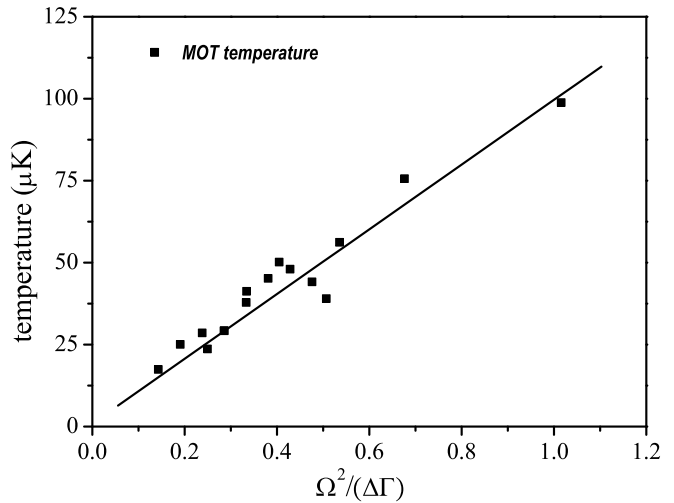}
\caption[Temperature of trapped atoms in a MOT in a standard beam configuration.]
{Temperatures of trapped atoms in a MOT in a standard beam configuration
are shown as a function of the light-shift parameter,
$\Omega^2/(|\varDelta|\Gamma)$.
The solid squares are measured temperatures and the solid line is a linear fit.
The detuning is varied from $-1.2\Gamma$ to $-3.7\Gamma$.
The observed minimum temperature is about 20 $\mu$K at the largest detuning of $\varDelta =-3.7\Gamma$.
}
\label{fig:typical_temperature}
\end{figure}

Figure \ref{fig:spectrum_vs_detuning}(b) shows time-averaged spectra measured 
at various detunings.
The spectral shapes with a narrow peak at the center and a 
Doppler-broadened pedestal are similar to those of the phase-stabilized 
MOT with high atomic densities and also similar to the spectra firstly observed 
in the 3D optical molasses~\cite{PhysRevLett.65.33}.
The linewidth of the Doppler-broadened pedestal increases from 
200 kHz to 500 kHz as the detuning is decreased.

The atomic temperatures can be extracted from the widths 
of the Doppler-broadened spectra by ignoring the narrow peak at the center.
In the simple one-dimensional sub-Doppler cooling theory~\cite{Dalibard:89, Chang200265}, 
the temperature of sub-Doppler cooled atoms in a MOT is 
proportional to the trap intensity and inversely proportional to the detuning, 
$T_s \propto \hbar \Omega^2/k_B|\varDelta|$, where $\Omega = \sqrt{(I_0/2I_s)}\Gamma$ is 
the Rabi frequency per single trapping beam.
The measured temperatures for various trap parameters shown 
in Fig.~\ref{fig:typical_temperature} are consistent with the sub-Doppler cooling 
theory and also show a reasonable agreement with earlier measurements of the MOT temperature.
The Doppler-cooling limit temperature for a rubidium atom is 
$T_D = \hbar \Gamma / 2k_\mathrm{B} \sim 140~\mu K$,
which corresponds to spectral broadening of $\sim$600 kHz.
The measured temperatures are well below the Doppler-cooling limit $T_D$ as expected.

\section{Summary}
\label{sec:conclusion}
We have built a high-speed precision spectrum measurement system for a small number of 
cold atoms based on the heterodyne photon-counting-based second-order correlation spectroscopy.
Fluorescence spectra of the trapped atoms in a phase-stabilized optical-lattice-like MOT 
have been measured under various trap conditions.
The optical lattice created by trapping lasers of the MOT itself forms an array structure of chained 
anisotropic double-well potentials in 3D space.
A narrow Rayleigh peak and well-resolved Raman sidebands which are observed in fluorescence spectra 
show that the MOT behaves as a stable 3D optical lattices simultaneously and the atoms can be well 
confined in the sub-wavelength-size double-well potentials.
We have also shown that the linewidths of the fluorescence spectra can be affected significantly 
by the radiation trapping even in a dilute atomic gas ($n\sim10^{8\sim9}$/cm$^3$).
The increase in the atomic temperature due to the radiation trapping 
gives rise to the excitation of atoms in bound motional states of the shallow double-well potentials 
to upper continuum states. 
This results in the line broadening in the fluorescence spectrum.
With a high atomic density, the radiation trapping as well as 
the intra-trap and background gas collisions induce more substantial spectral broadenings.
The rate of light-induced intra-trap cold collisions in an optical-lattice-like MOT is enhanced 
with high atomic densities ($n>10^{10}$/cm$^3$) because of the reduction of the effective trap volume. 
However, the cold collisions are expected to be suppressed dramatically for low atomic temperature with relatively 
low atomic densities in the range of $10^8\sim10^9$/cm$^3$.

\begin{acknowledgments}
The authors acknowledge fruitful discussions with Jai-Hyung Lee and Changwon Park.

\end{acknowledgments}


\begin{thebibliography}{30}
\expandafter\ifx\csname natexlab\endcsname\relax\def\natexlab#1{#1}\fi
\expandafter\ifx\csname bibnamefont\endcsname\relax
  \def\bibnamefont#1{#1}\fi
\expandafter\ifx\csname bibfnamefont\endcsname\relax
  \def\bibfnamefont#1{#1}\fi
\expandafter\ifx\csname citenamefont\endcsname\relax
  \def\citenamefont#1{#1}\fi
\expandafter\ifx\csname url\endcsname\relax
  \def\url#1{\texttt{#1}}\fi
\expandafter\ifx\csname urlprefix\endcsname\relax\def\urlprefix{URL }\fi
\providecommand{\bibinfo}[2]{#2}
\providecommand{\eprint}[2][]{\url{#2}}

\bibitem[{\citenamefont{Raab et~al.}(1987)\citenamefont{Raab, Prentiss, Cable,
  Chu, and Pritchard}}]{PhysRevLett.59.2631}
\bibinfo{author}{\bibfnamefont{E.~L.} \bibnamefont{Raab}},
  \bibinfo{author}{\bibfnamefont{M.}~\bibnamefont{Prentiss}},
  \bibinfo{author}{\bibfnamefont{A.}~\bibnamefont{Cable}},
  \bibinfo{author}{\bibfnamefont{S.}~\bibnamefont{Chu}}, \bibnamefont{and}
  \bibinfo{author}{\bibfnamefont{D.~E.} \bibnamefont{Pritchard}},
  \bibinfo{journal}{Phys. Rev. Lett.} \textbf{\bibinfo{volume}{59}},
  \bibinfo{pages}{2631} (\bibinfo{year}{1987}).

\bibitem[{\citenamefont{Westbrook et~al.}(1990)\citenamefont{Westbrook, Watts,
  Tanner, Rolston, Phillips, Lett, and Gould}}]{PhysRevLett.65.33}
\bibinfo{author}{\bibfnamefont{C.~I.} \bibnamefont{Westbrook}},
  \bibinfo{author}{\bibfnamefont{R.~N.} \bibnamefont{Watts}},
  \bibinfo{author}{\bibfnamefont{C.~E.} \bibnamefont{Tanner}},
  \bibinfo{author}{\bibfnamefont{S.~L.} \bibnamefont{Rolston}},
  \bibinfo{author}{\bibfnamefont{W.~D.} \bibnamefont{Phillips}},
  \bibinfo{author}{\bibfnamefont{P.~D.} \bibnamefont{Lett}}, \bibnamefont{and}
  \bibinfo{author}{\bibfnamefont{P.~L.} \bibnamefont{Gould}},
  \bibinfo{journal}{Phys. Rev. Lett.} \textbf{\bibinfo{volume}{65}},
  \bibinfo{pages}{33} (\bibinfo{year}{1990}).

\bibitem[{\citenamefont{Jessen et~al.}(1992)\citenamefont{Jessen, Gerz, Lett,
  Phillips, Rolston, Spreeuw, and Westbrook}}]{PhysRevLett.69.49}
\bibinfo{author}{\bibfnamefont{P.~S.} \bibnamefont{Jessen}},
  \bibinfo{author}{\bibfnamefont{C.}~\bibnamefont{Gerz}},
  \bibinfo{author}{\bibfnamefont{P.~D.} \bibnamefont{Lett}},
  \bibinfo{author}{\bibfnamefont{W.~D.} \bibnamefont{Phillips}},
  \bibinfo{author}{\bibfnamefont{S.~L.} \bibnamefont{Rolston}},
  \bibinfo{author}{\bibfnamefont{R.~J.~C.} \bibnamefont{Spreeuw}},
  \bibnamefont{and} \bibinfo{author}{\bibfnamefont{C.~I.}
  \bibnamefont{Westbrook}}, \bibinfo{journal}{Phys. Rev. Lett.}
  \textbf{\bibinfo{volume}{69}}, \bibinfo{pages}{49} (\bibinfo{year}{1992}).

\bibitem[{\citenamefont{Hemmerich and H\"ansch}(1993)}]{PhysRevLett.70.410}
\bibinfo{author}{\bibfnamefont{A.}~\bibnamefont{Hemmerich}} \bibnamefont{and}
  \bibinfo{author}{\bibfnamefont{T.~W.} \bibnamefont{H\"ansch}},
  \bibinfo{journal}{Phys. Rev. Lett.} \textbf{\bibinfo{volume}{70}},
  \bibinfo{pages}{410} (\bibinfo{year}{1993}).

\bibitem[{\citenamefont{Grynberg et~al.}(1993)\citenamefont{Grynberg, Lounis,
  Verkerk, Courtois, and Salomon}}]{Grynberg_PRL_1993}
\bibinfo{author}{\bibfnamefont{G.}~\bibnamefont{Grynberg}},
  \bibinfo{author}{\bibfnamefont{B.}~\bibnamefont{Lounis}},
  \bibinfo{author}{\bibfnamefont{P.}~\bibnamefont{Verkerk}},
  \bibinfo{author}{\bibfnamefont{J.-Y.} \bibnamefont{Courtois}},
  \bibnamefont{and} \bibinfo{author}{\bibfnamefont{C.}~\bibnamefont{Salomon}},
  \bibinfo{journal}{Phys. Rev. Lett.} \textbf{\bibinfo{volume}{70}},
  \bibinfo{pages}{2249} (\bibinfo{year}{1993}).

\bibitem[{\citenamefont{Gatzke et~al.}(1997)\citenamefont{Gatzke, Birkl,
  Jessen, Kastberg, Rolston, and Phillips}}]{PhysRevA.55.R3987}
\bibinfo{author}{\bibfnamefont{M.}~\bibnamefont{Gatzke}},
  \bibinfo{author}{\bibfnamefont{G.}~\bibnamefont{Birkl}},
  \bibinfo{author}{\bibfnamefont{P.~S.} \bibnamefont{Jessen}},
  \bibinfo{author}{\bibfnamefont{A.}~\bibnamefont{Kastberg}},
  \bibinfo{author}{\bibfnamefont{S.~L.} \bibnamefont{Rolston}},
  \bibnamefont{and} \bibinfo{author}{\bibfnamefont{W.~D.}
  \bibnamefont{Phillips}}, \bibinfo{journal}{Phys. Rev. A}
  \textbf{\bibinfo{volume}{55}}, \bibinfo{pages}{R3987} (\bibinfo{year}{1997}).

\bibitem[{\citenamefont{Schadwinkel et~al.}(1999)\citenamefont{Schadwinkel,
  Reiter, Gomer, and Meschede}}]{PhysRevA.61.013409}
\bibinfo{author}{\bibfnamefont{H.}~\bibnamefont{Schadwinkel}},
  \bibinfo{author}{\bibfnamefont{U.}~\bibnamefont{Reiter}},
  \bibinfo{author}{\bibfnamefont{V.}~\bibnamefont{Gomer}}, \bibnamefont{and}
  \bibinfo{author}{\bibfnamefont{D.}~\bibnamefont{Meschede}},
  \bibinfo{journal}{Phys. Rev. A} \textbf{\bibinfo{volume}{61}},
  \bibinfo{pages}{013409} (\bibinfo{year}{1999}).

\bibitem[{\citenamefont{Rauschenbeutel
  et~al.}(1998)\citenamefont{Rauschenbeutel, Schadwinkel, Gomer, and
  Meschede}}]{OptComm.148.45}
\bibinfo{author}{\bibfnamefont{A.}~\bibnamefont{Rauschenbeutel}},
  \bibinfo{author}{\bibfnamefont{H.}~\bibnamefont{Schadwinkel}},
  \bibinfo{author}{\bibfnamefont{V.}~\bibnamefont{Gomer}}, \bibnamefont{and}
  \bibinfo{author}{\bibfnamefont{D.}~\bibnamefont{Meschede}},
  \bibinfo{journal}{Opt. Comm.} \textbf{\bibinfo{volume}{148}},
  \bibinfo{pages}{45} (\bibinfo{year}{1998}).

\bibitem[{\citenamefont{Dalibard and Cohen-Tannoudji}(1989)}]{Dalibard:89}
\bibinfo{author}{\bibfnamefont{J.}~\bibnamefont{Dalibard}} \bibnamefont{and}
  \bibinfo{author}{\bibfnamefont{C.}~\bibnamefont{Cohen-Tannoudji}},
  \bibinfo{journal}{J. Opt. Soc. Am. B} \textbf{\bibinfo{volume}{6}},
  \bibinfo{pages}{2023} (\bibinfo{year}{1989}).

\bibitem[{\citenamefont{Weiner et~al.}(1999)\citenamefont{Weiner, Bagnato,
  Zilio, and Julienne}}]{cold_collision1}
\bibinfo{author}{\bibfnamefont{J.}~\bibnamefont{Weiner}},
  \bibinfo{author}{\bibfnamefont{V.~S.} \bibnamefont{Bagnato}},
  \bibinfo{author}{\bibfnamefont{S.}~\bibnamefont{Zilio}}, \bibnamefont{and}
  \bibinfo{author}{\bibfnamefont{P.~S.} \bibnamefont{Julienne}},
  \bibinfo{journal}{Rev. Mod. Phys.} \textbf{\bibinfo{volume}{71}},
  \bibinfo{pages}{1} (\bibinfo{year}{1999}).

\bibitem[{\citenamefont{Ueberholz et~al.}(2000)\citenamefont{Ueberholz, Kuhr,
  Frese, Meschede, and Gomer}}]{cold_collision2}
\bibinfo{author}{\bibfnamefont{B.}~\bibnamefont{Ueberholz}},
  \bibinfo{author}{\bibfnamefont{S.}~\bibnamefont{Kuhr}},
  \bibinfo{author}{\bibfnamefont{D.}~\bibnamefont{Frese}},
  \bibinfo{author}{\bibfnamefont{D.}~\bibnamefont{Meschede}}, \bibnamefont{and}
  \bibinfo{author}{\bibfnamefont{V.}~\bibnamefont{Gomer}}, \bibinfo{journal}{J.
  Phys. B: At. Mol. Opt. Phys.} \textbf{\bibinfo{volume}{33}},
  \bibinfo{pages}{L135} (\bibinfo{year}{2000}).

\bibitem[{\citenamefont{Choi et~al.}(2007)\citenamefont{Choi, Yoon, Kang, Kim,
  Lee, and An}}]{cold_collision3}
\bibinfo{author}{\bibfnamefont{Y.}~\bibnamefont{Choi}},
  \bibinfo{author}{\bibfnamefont{S.}~\bibnamefont{Yoon}},
  \bibinfo{author}{\bibfnamefont{S.}~\bibnamefont{Kang}},
  \bibinfo{author}{\bibfnamefont{W.}~\bibnamefont{Kim}},
  \bibinfo{author}{\bibfnamefont{J.-H.} \bibnamefont{Lee}}, \bibnamefont{and}
  \bibinfo{author}{\bibfnamefont{K.}~\bibnamefont{An}}, \bibinfo{journal}{Phys.
  Rev. A} \textbf{\bibinfo{volume}{76}}, \bibinfo{pages}{013402}
  (\bibinfo{year}{2007}).

\bibitem[{\citenamefont{Lawall et~al.}(1998)\citenamefont{Lawall, Orzel, and
  Rolston}}]{PhysRevLett.80.480}
\bibinfo{author}{\bibfnamefont{J.}~\bibnamefont{Lawall}},
  \bibinfo{author}{\bibfnamefont{C.}~\bibnamefont{Orzel}}, \bibnamefont{and}
  \bibinfo{author}{\bibfnamefont{S.~L.} \bibnamefont{Rolston}},
  \bibinfo{journal}{Phys. Rev. Lett.} \textbf{\bibinfo{volume}{80}},
  \bibinfo{pages}{480} (\bibinfo{year}{1998}).


\bibitem[{\citenamefont{Grison et~al.}(1991)\citenamefont{Grison, Lounis,
  Salomon, Courtois, and Grynberg}}]{Europhyslett.15.149}
\bibinfo{author}{\bibfnamefont{D.}~\bibnamefont{Grison}},
  \bibinfo{author}{\bibfnamefont{B.}~\bibnamefont{Lounis}},
  \bibinfo{author}{\bibfnamefont{C.}~\bibnamefont{Salomon}},
  \bibinfo{author}{\bibfnamefont{J.~Y.} \bibnamefont{Courtois}},
  \bibnamefont{and} \bibinfo{author}{\bibfnamefont{G.}~\bibnamefont{Grynberg}},
  \bibinfo{journal}{Europhys. Lett.} \textbf{\bibinfo{volume}{15}},
  \bibinfo{pages}{149} (\bibinfo{year}{1991}).

\bibitem[{\citenamefont{Tabosa et~al.}(1991)\citenamefont{Tabosa, Chen, Hu,
  Lee, and Kimble}}]{PhysRevLett.66.3245}
\bibinfo{author}{\bibfnamefont{J.~W.~R.} \bibnamefont{Tabosa}},
  \bibinfo{author}{\bibfnamefont{G.}~\bibnamefont{Chen}},
  \bibinfo{author}{\bibfnamefont{Z.}~\bibnamefont{Hu}},
  \bibinfo{author}{\bibfnamefont{R.~B.} \bibnamefont{Lee}}, \bibnamefont{and}
  \bibinfo{author}{\bibfnamefont{H.~J.} \bibnamefont{Kimble}},
  \bibinfo{journal}{Phys. Rev. Lett.} \textbf{\bibinfo{volume}{66}},
  \bibinfo{pages}{3245} (\bibinfo{year}{1991}).

\bibitem[{\citenamefont{Brzozowski et~al.}(2005)\citenamefont{Brzozowski,
  Brzozowska, Zachorowski, Zawada, and Gawlik}}]{PhysRevA.71.013401}
\bibinfo{author}{\bibfnamefont{T.~M.} \bibnamefont{Brzozowski}},
  \bibinfo{author}{\bibfnamefont{M.}~\bibnamefont{Brzozowska}},
  \bibinfo{author}{\bibfnamefont{J.}~\bibnamefont{Zachorowski}},
  \bibinfo{author}{\bibfnamefont{M.}~\bibnamefont{Zawada}}, \bibnamefont{and}
  \bibinfo{author}{\bibfnamefont{W.}~\bibnamefont{Gawlik}},
  \bibinfo{journal}{Phys. Rev. A} \textbf{\bibinfo{volume}{71}},
  \bibinfo{pages}{013401} (\bibinfo{year}{2005}).

\bibitem[{\citenamefont{Souther et~al.}(2010)\citenamefont{Souther, Wagner,
  Harnish, Briel, and Bali}}]{LaserPhysLett.7.321}
\bibinfo{author}{\bibfnamefont{N.}~\bibnamefont{Souther}},
  \bibinfo{author}{\bibfnamefont{R.}~\bibnamefont{Wagner}},
  \bibinfo{author}{\bibfnamefont{P.}~\bibnamefont{Harnish}},
  \bibinfo{author}{\bibfnamefont{M.}~\bibnamefont{Briel}}, \bibnamefont{and}
  \bibinfo{author}{\bibfnamefont{S.}~\bibnamefont{Bali}},
  \bibinfo{journal}{Laser Physics Letters} \textbf{\bibinfo{volume}{7}},
  \bibinfo{pages}{321} (\bibinfo{year}{2010}).


\bibitem[{\citenamefont{Bali et~al.}(1996)\citenamefont{Bali, Hoffmann,
  Sim\'an, and Walker}}]{PhysRevA.53.3469}
\bibinfo{author}{\bibfnamefont{S.}~\bibnamefont{Bali}},
  \bibinfo{author}{\bibfnamefont{D.}~\bibnamefont{Hoffmann}},
  \bibinfo{author}{\bibfnamefont{J.}~\bibnamefont{Sim\'an}}, \bibnamefont{and}
  \bibinfo{author}{\bibfnamefont{T.}~\bibnamefont{Walker}},
  \bibinfo{journal}{Phys. Rev. A} \textbf{\bibinfo{volume}{53}},
  \bibinfo{pages}{3469} (\bibinfo{year}{1996}).

\bibitem[{\citenamefont{Beeler et~al.}(2003)\citenamefont{Beeler, Stites, Kim,
  Feeney, and Bali}}]{PhysRevA.68.013411}
\bibinfo{author}{\bibfnamefont{M.}~\bibnamefont{Beeler}},
  \bibinfo{author}{\bibfnamefont{R.}~\bibnamefont{Stites}},
  \bibinfo{author}{\bibfnamefont{S.}~\bibnamefont{Kim}},
  \bibinfo{author}{\bibfnamefont{L.}~\bibnamefont{Feeney}}, \bibnamefont{and}
  \bibinfo{author}{\bibfnamefont{S.}~\bibnamefont{Bali}},
  \bibinfo{journal}{Phys. Rev. A} \textbf{\bibinfo{volume}{68}},
  \bibinfo{pages}{013411} (\bibinfo{year}{2003}).

\bibitem[{\citenamefont{Nakayama et~al.}(2010)\citenamefont{Nakayama,
  Yoshikawa, Matsumoto, Torii, and Kuga}}]{Nakayama:10}
\bibinfo{author}{\bibfnamefont{K.}~\bibnamefont{Nakayama}},
  \bibinfo{author}{\bibfnamefont{Y.}~\bibnamefont{Yoshikawa}},
  \bibinfo{author}{\bibfnamefont{H.}~\bibnamefont{Matsumoto}},
  \bibinfo{author}{\bibfnamefont{Y.}~\bibnamefont{Torii}}, \bibnamefont{and}
  \bibinfo{author}{\bibfnamefont{T.}~\bibnamefont{Kuga}},
  \bibinfo{journal}{Opt. Express} \textbf{\bibinfo{volume}{18}},
  \bibinfo{pages}{6604} (\bibinfo{year}{2010}).

\bibitem[{\citenamefont{Kim et~al.}(2011)\citenamefont{Kim, Park, Kim, Choi,
  Kang, Lim, Lee, Ihm, and An}}]{NanoLett.11.729}
\bibinfo{author}{\bibfnamefont{W.}~\bibnamefont{Kim}},
  \bibinfo{author}{\bibfnamefont{C.}~\bibnamefont{Park}},
  \bibinfo{author}{\bibfnamefont{J.-R.} \bibnamefont{Kim}},
  \bibinfo{author}{\bibfnamefont{Y.}~\bibnamefont{Choi}},
  \bibinfo{author}{\bibfnamefont{S.}~\bibnamefont{Kang}},
  \bibinfo{author}{\bibfnamefont{S.}~\bibnamefont{Lim}},
  \bibinfo{author}{\bibfnamefont{Y.-L.} \bibnamefont{Lee}},
  \bibinfo{author}{\bibfnamefont{J.}~\bibnamefont{Ihm}}, \bibnamefont{and}
  \bibinfo{author}{\bibfnamefont{K.}~\bibnamefont{An}}, \bibinfo{journal}{Nano
  Letters} \textbf{\bibinfo{volume}{11}}, \bibinfo{pages}{729}
  (\bibinfo{year}{2011}).

\bibitem[{\citenamefont{Hong et~al.}(2006)\citenamefont{Hong, Seo, Lee, Choi,
  Lee, and An}}]{Hong:06}
\bibinfo{author}{\bibfnamefont{H.-G.} \bibnamefont{Hong}},
  \bibinfo{author}{\bibfnamefont{W.}~\bibnamefont{Seo}},
  \bibinfo{author}{\bibfnamefont{M.}~\bibnamefont{Lee}},
  \bibinfo{author}{\bibfnamefont{W.}~\bibnamefont{Choi}},
  \bibinfo{author}{\bibfnamefont{J.-H.} \bibnamefont{Lee}}, \bibnamefont{and}
  \bibinfo{author}{\bibfnamefont{K.}~\bibnamefont{An}}, \bibinfo{journal}{Opt.
  Lett.} \textbf{\bibinfo{volume}{31}}, \bibinfo{pages}{3182}
  (\bibinfo{year}{2006}).

\bibitem[{\citenamefont{Dicke}(1953)}]{Dicke}
\bibinfo{author}{\bibfnamefont{R.~H.} \bibnamefont{Dicke}},
  \bibinfo{journal}{Phys. Rev.} \textbf{\bibinfo{volume}{89}},
  \bibinfo{pages}{472} (\bibinfo{year}{1953}).

\bibitem[{\citenamefont{Das et~al.}(2010)\citenamefont{Das, Shirasaki, Nayak,
  Morinaga, Kien, and Hakuta}}]{Das:10}
\bibinfo{author}{\bibfnamefont{M.}~\bibnamefont{Das}},
  \bibinfo{author}{\bibfnamefont{A.}~\bibnamefont{Shirasaki}},
  \bibinfo{author}{\bibfnamefont{K.~P.} \bibnamefont{Nayak}},
  \bibinfo{author}{\bibfnamefont{M.}~\bibnamefont{Morinaga}},
  \bibinfo{author}{\bibfnamefont{F.~L.} \bibnamefont{Kien}}, \bibnamefont{and}
  \bibinfo{author}{\bibfnamefont{K.}~\bibnamefont{Hakuta}},
  \bibinfo{journal}{Opt. Express} \textbf{\bibinfo{volume}{18}},
  \bibinfo{pages}{17154} (\bibinfo{year}{2010}).
  
\bibitem[{\citenamefont{Walker et~al.}(1990)\citenamefont{Walker, Sesko, and
  Wieman}}]{radiation_trapping1}
\bibinfo{author}{\bibfnamefont{T.}~\bibnamefont{Walker}},
  \bibinfo{author}{\bibfnamefont{D.}~\bibnamefont{Sesko}}, \bibnamefont{and}
  \bibinfo{author}{\bibfnamefont{C.}~\bibnamefont{Wieman}},
  \bibinfo{journal}{Phys. Rev. Lett.} \textbf{\bibinfo{volume}{64}},
  \bibinfo{pages}{408} (\bibinfo{year}{1990}).

\bibitem[{\citenamefont{Kim et~al.}(2010)\citenamefont{Kim, Park, Kim, Lee,
  Ihm, and An}}]{arXiv}
\bibinfo{author}{\bibfnamefont{W.}~\bibnamefont{Kim}},
  \bibinfo{author}{\bibfnamefont{C.}~\bibnamefont{Park}},
  \bibinfo{author}{\bibfnamefont{J.-R.} \bibnamefont{Kim}},
  \bibinfo{author}{\bibfnamefont{Y.-L.} \bibnamefont{Lee}},
  \bibinfo{author}{\bibfnamefont{J.}~\bibnamefont{Ihm}}, \bibnamefont{and}
  \bibinfo{author}{\bibfnamefont{K.}~\bibnamefont{An}},
  \bibinfo{journal}{arXiv:1010.6023v2}  (\bibinfo{year}{2010}).

\bibitem[{\citenamefont{Yoon et~al.}(2007)\citenamefont{Yoon, Choi, Park, Ji,
  Lee, and An}}]{yoon}
\bibinfo{author}{\bibfnamefont{S.}~\bibnamefont{Yoon}},
  \bibinfo{author}{\bibfnamefont{Y.}~\bibnamefont{Choi}},
  \bibinfo{author}{\bibfnamefont{S.}~\bibnamefont{Park}},
  \bibinfo{author}{\bibfnamefont{W.}~\bibnamefont{Ji}},
  \bibinfo{author}{\bibfnamefont{J.-H.} \bibnamefont{Lee}}, \bibnamefont{and}
  \bibinfo{author}{\bibfnamefont{K.}~\bibnamefont{An}},
  \bibinfo{journal}{Journal of Physics: Conference Series}
  \textbf{\bibinfo{volume}{80}}, \bibinfo{pages}{012046}
  (\bibinfo{year}{2007}).

\bibitem[{\citenamefont{Courtois and Grynberg}(1992)}]{PhysRevA.46.7060}
\bibinfo{author}{\bibfnamefont{J.-Y.} \bibnamefont{Courtois}} \bibnamefont{and}
  \bibinfo{author}{\bibfnamefont{G.}~\bibnamefont{Grynberg}},
  \bibinfo{journal}{Phys. Rev. A} \textbf{\bibinfo{volume}{46}},
  \bibinfo{pages}{7060} (\bibinfo{year}{1992}).

\bibitem[{\citenamefont{Morsch et~al.}(2000)\citenamefont{Morsch, Jones, and
  Meacher}}]{PhysRevA.61.023410}
\bibinfo{author}{\bibfnamefont{O.}~\bibnamefont{Morsch}},
  \bibinfo{author}{\bibfnamefont{P.~H.} \bibnamefont{Jones}}, \bibnamefont{and}
  \bibinfo{author}{\bibfnamefont{D.~R.} \bibnamefont{Meacher}},
  \bibinfo{journal}{Phys. Rev. A} \textbf{\bibinfo{volume}{61}},
  \bibinfo{pages}{023410} (\bibinfo{year}{2000}).

\bibitem[{\citenamefont{Chang and Minogin}(2002)}]{Chang200265}
\bibinfo{author}{\bibfnamefont{S.}~\bibnamefont{Chang}} \bibnamefont{and}
  \bibinfo{author}{\bibfnamefont{V.}~\bibnamefont{Minogin}},
  \bibinfo{journal}{Physics Reports} \textbf{\bibinfo{volume}{365}},
  \bibinfo{pages}{65} (\bibinfo{year}{2002}).

\end{thebibliography}

\end{document}